\documentclass[twoside]{ilcws07}
\usepackage[latin1]{inputenc}
\usepackage[dvips]{graphicx,epsfig,color}
\usepackage{wrapfig,rotating}
\usepackage{amssymb,amsmath,array}

\begin{document}
\title{Test Stand Measurements for an ILC Polarimeter}
\author{Daniela K\"afer$^1$, 
  Oleg Eyser$^1$, 
  Christian Helebrant$^{1,2}$, 
  Jenny List$^1$, 
  Ulrich Velte$^3$
%
\vspace{.3cm}\\
1- Deutsches Elektronen Synchrotron (DESY) - Hamburg, Germany \\[0.4mm]
2- Universit\"at Hamburg, Germany          \\[0.4mm]
3- Leibniz Universit\"at Hannover, Germany \\[-2.4mm] 
}

\maketitle

\begin{abstract}
  The setup of two different small scale teststands for measurements regarding 
  an electron Cherenkov detector as part of the ILC polarimeters is presented. 
  Component measurements already carried out are analyzed and others, foreseen 
  for the near future, are discussed. The larger one of the two teststands 
  features the old Cherenkov detector of the SLD experiment, which will be used 
  as a reference for a number of crucial measurements. 
  Especially, the requirements for the non-linearity of the read-out chain are 
  studied in greater detail and methods for its precise measurement before and 
  during operation are being developed accordingly. 
\end{abstract}

\section{Polarisation: Measurement Principle}
At the ILC it will be necessary to measure the beam polarisation with a precision 
of 0.25\% to fully exploit the physics potential of machine and detectors~\cite{bib:PhysicsCase}. 
These measurements will be realised via Compton polarimeters, where Cherenkov counters 
detect the backscattered Compton electrons. The Compton cross sections for different 
configurations of the laser light 
\begin{wrapfigure}{l}{0.70\textwidth}
  \setlength{\unitlength}{1.0cm}
  \begin{picture}(10.4, 8.47)
    \put( 0.0, 3.8)  {\epsfig{file=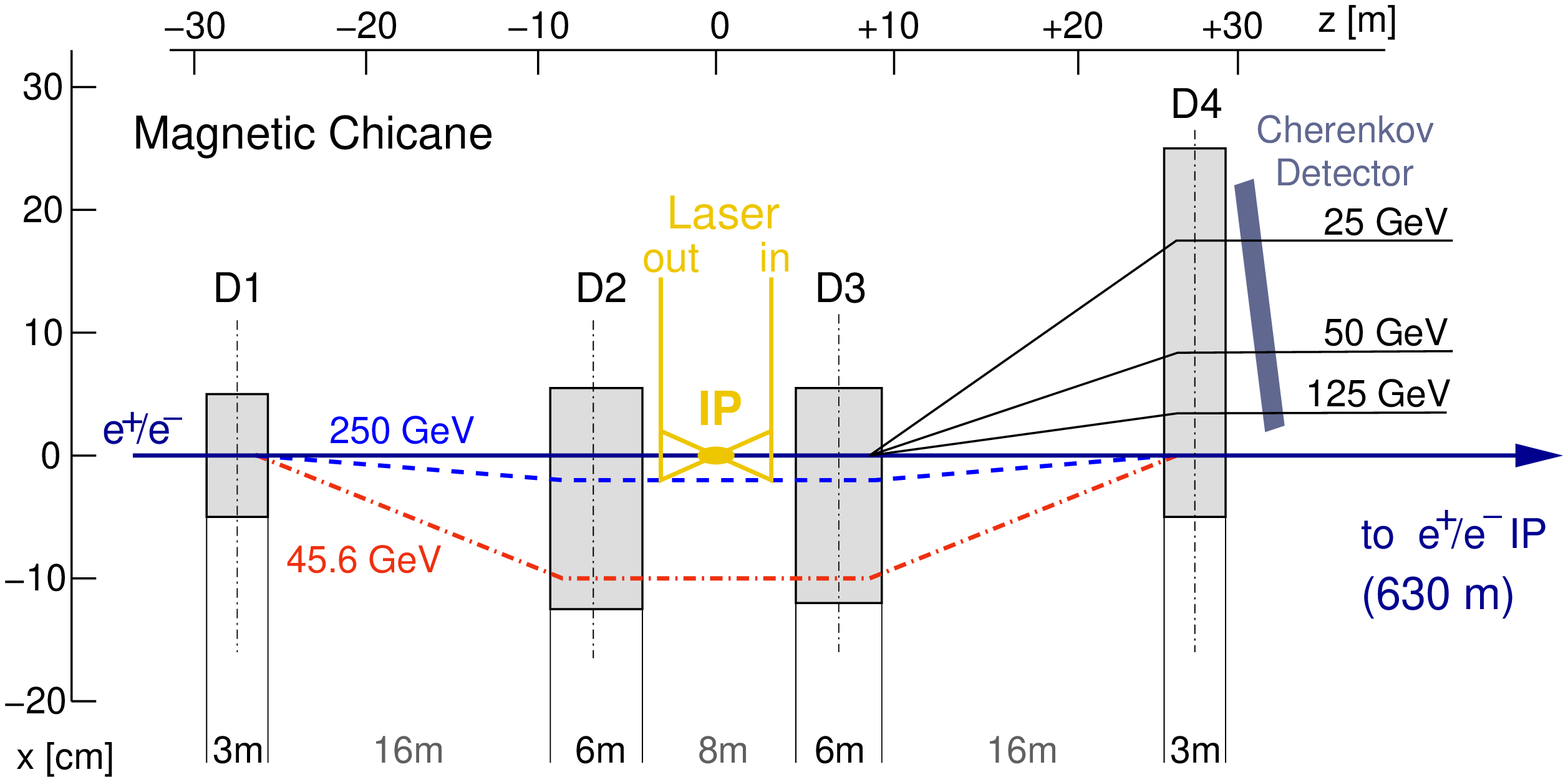, bb= 0 0 730 370, clip= , width=1.00\linewidth}}
    \put( 0.0, 0.0)  {\epsfig{file=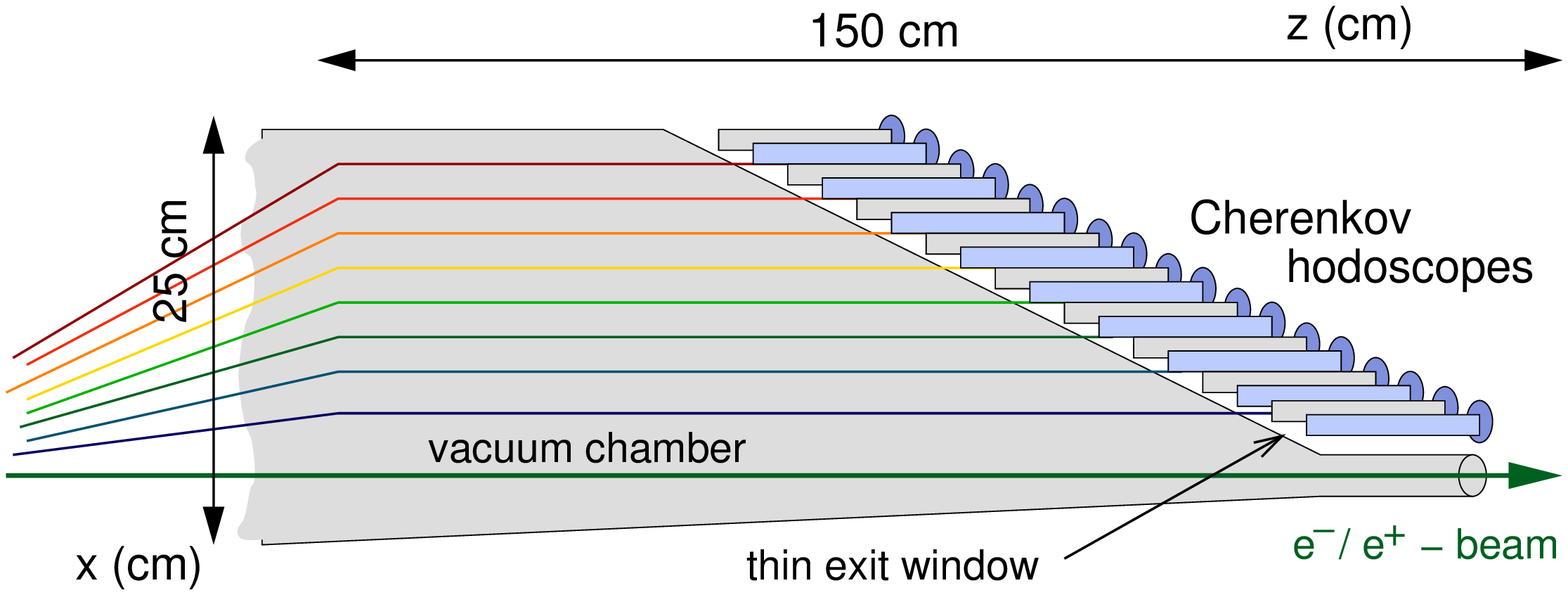, bb= 0 0 642 240, clip= , width=0.94\linewidth}}
    \put( 9.3, 8.3)  {(a)}
    \put( 9.3, 3.4)  {(b)}
  \end{picture}
  \vspace*{-5.0mm}
  \caption{(a) The magnetic spectrometer and (b) the layout of the 
    gas cherenkov hodoscope for the polarisation measurements.}
  \label{fig:MagnChic-CherHodoLayout}
\end{wrapfigure}
and $e^+/e^-$ spin helicities are different, allowing for an asymmetry 
measurement, from which the polarisation level of the $e^+/e^-$-beams 
can finally be determined. 
For the upstream polarimeter a special magnetic chicane 
(Fig.~\ref{fig:MagnChic-CherHodoLayout}(a)) is envisioned, so that the Compton 
edge of the backscattered electron beam will always be at the same position in the 
Cherenkov detector regardless of the energy of the original electron/positron beam. 
Circularly polarised laser light is scattered off the $e^+/e^-$-beams with about 
$10^3$ interactions per bunch. The scattered $e^+/e^-$ are then deflected by the 
dipole magnetic fields of 
the chicane and led to the detector, whose current baseline design~\cite{bib:Tesla} 
consists of gas tubes read out by photomultipliers (Fig.~\ref{fig:MagnChic-CherHodoLayout}(b)). 
The incident $e^+/e^-$ generate Cherenkov radiation inside the gas tubes, which is 
then detected by photomultipliers. However, alternative design possibilities, using 
quartz fibers and silicon photomultipliers, are also being studied.

For the detector R\&D, many different aspects have to be taken into account and 
optimised. Among these are not only the choice of gas or the inside mirror-coating 
of the Cherenkov gas tubes (currently similar to those used for the polarimeter of 
the SLD detector at SLAC), but also different aspects of the photomultiplier and 
read-out electronics. 
It will, for example, be necessary to optimise the quantum efficiency, the sensitive 
area and the dynamic range of the photomultipliers, but also the reflectivity and 
the light extraction from the gas tubes. 
Furthermore, since the goal of achieving a polarisation measurement with a precision 
of 0.25\% is very ambitious, the linearity of all detector components is extremely 
important. All non-linear effects (photodetectors, electronics, etc.) need to be 
measured precisely and corrected for if necessary.

\section{The Component Test Stand}
\begin{wrapfigure}{l}{0.48\textwidth}
  \setlength{\unitlength}{1.0cm}
  \begin{picture}(6.0, 1.84)
    \put( 0.0, 0.0)  {\epsfig{file=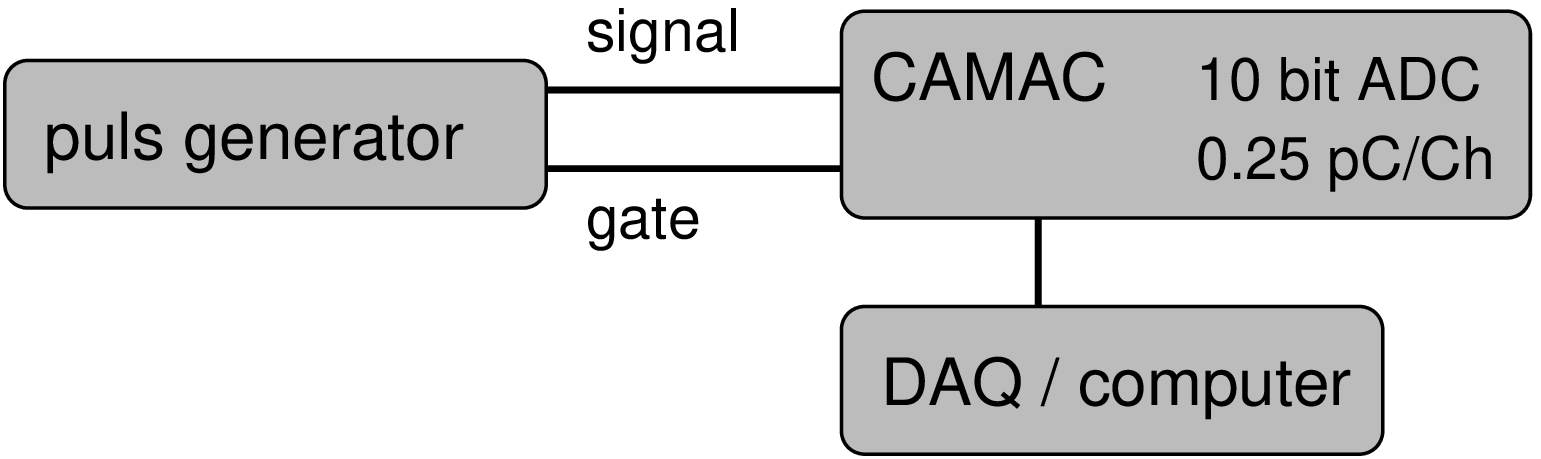, bb= 0 0 445 135, clip= , width=1.00\linewidth}}
  \end{picture}
  \vspace*{-2.0mm}
  \caption{The setup for the electronics tests.}
  \label{fig:ElectronicsTest}
  \vspace*{-5mm}
\end{wrapfigure}

\begin{figure}[!h]
  \setlength{\unitlength}{1.0cm}
  \begin{picture}(14.0, 4.4)
    \put( 0.0, 0.0)  {\epsfig{file=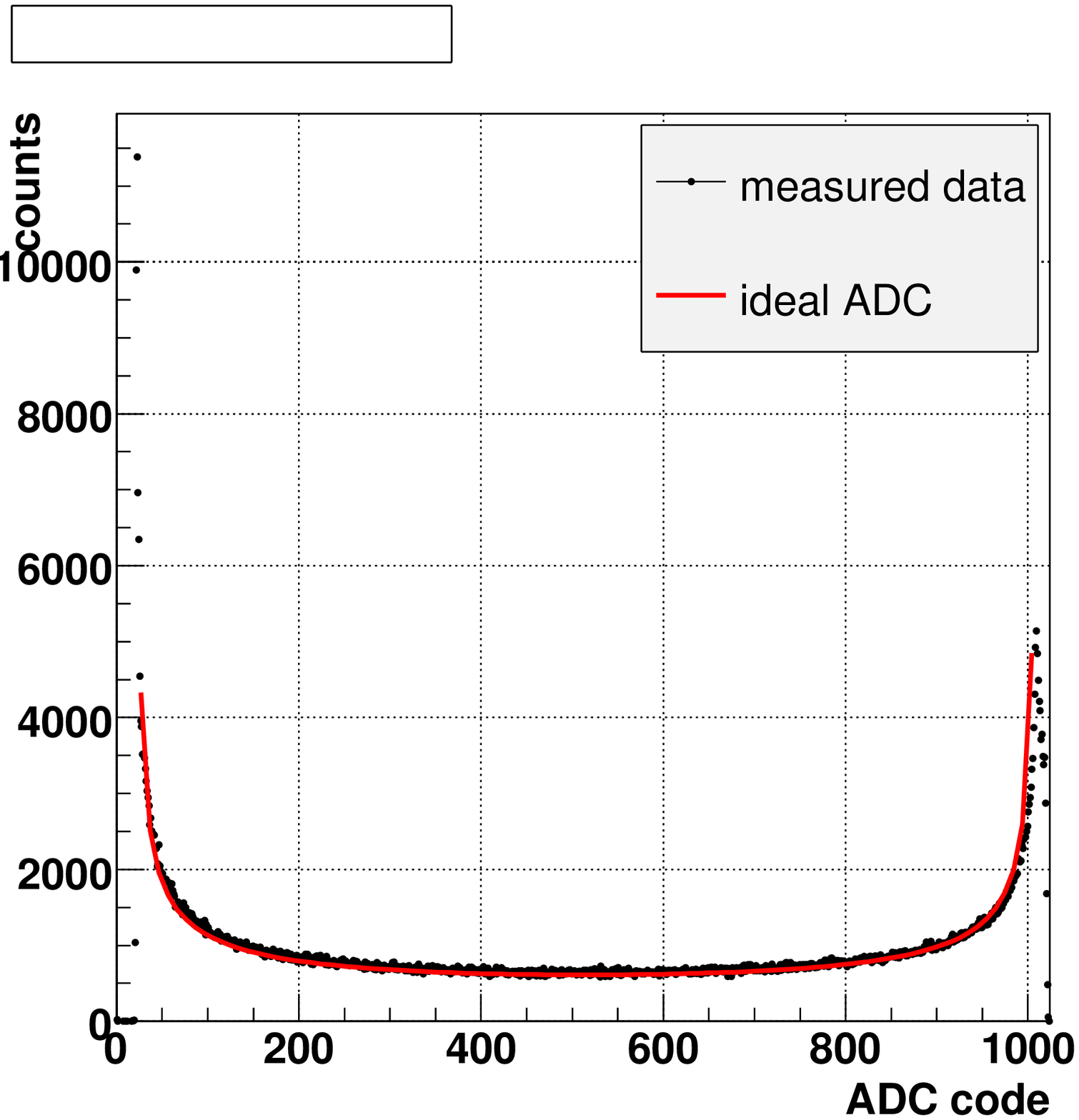, bb= 0 0 567 500, clip= , width=0.35\linewidth}}
    \put( 4.8, 0.0)  {\epsfig{file=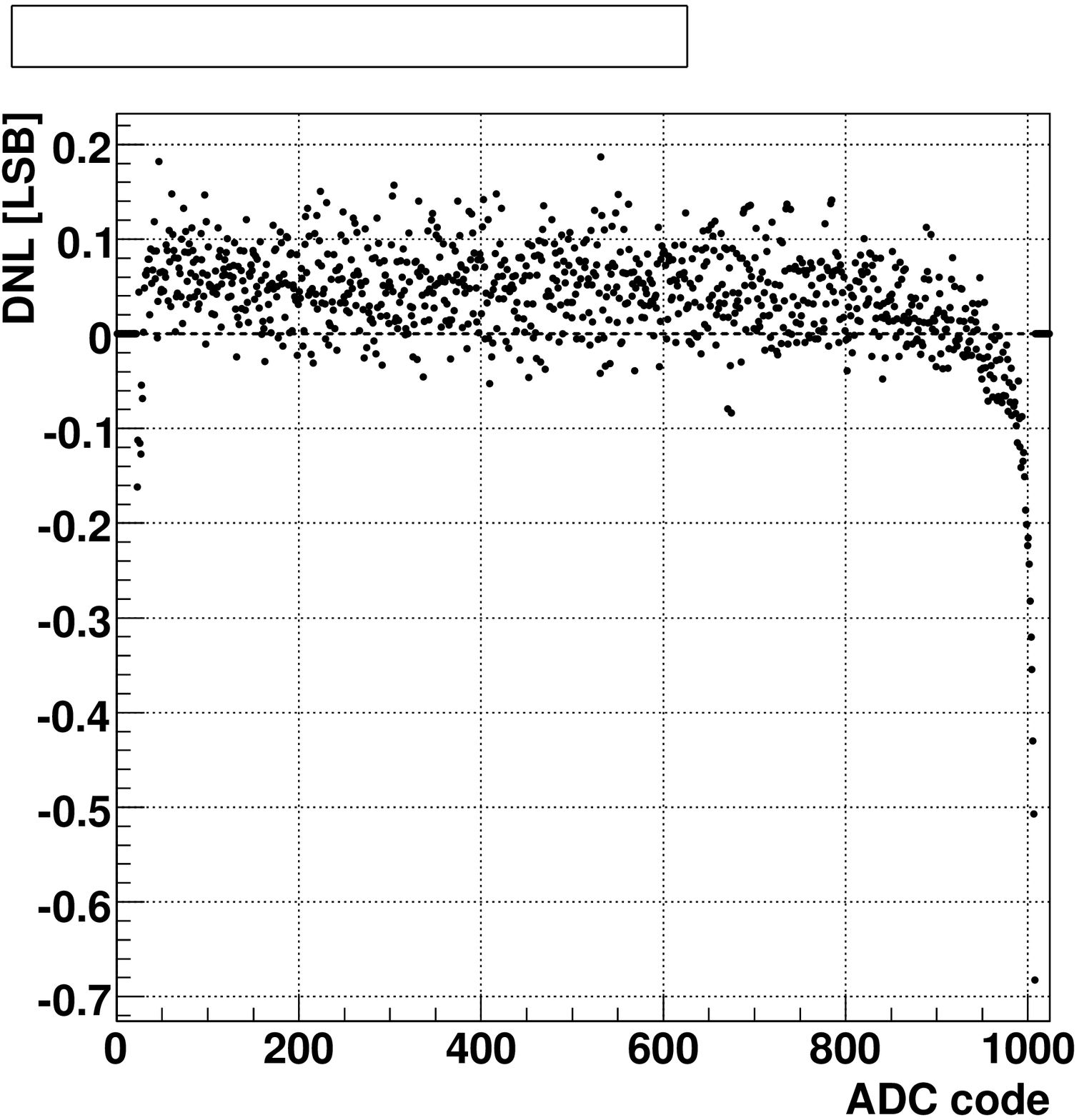, bb= 0 0 567 500, clip= , width=0.35\linewidth}}
    \put( 9.5, 0.0)  {\epsfig{file=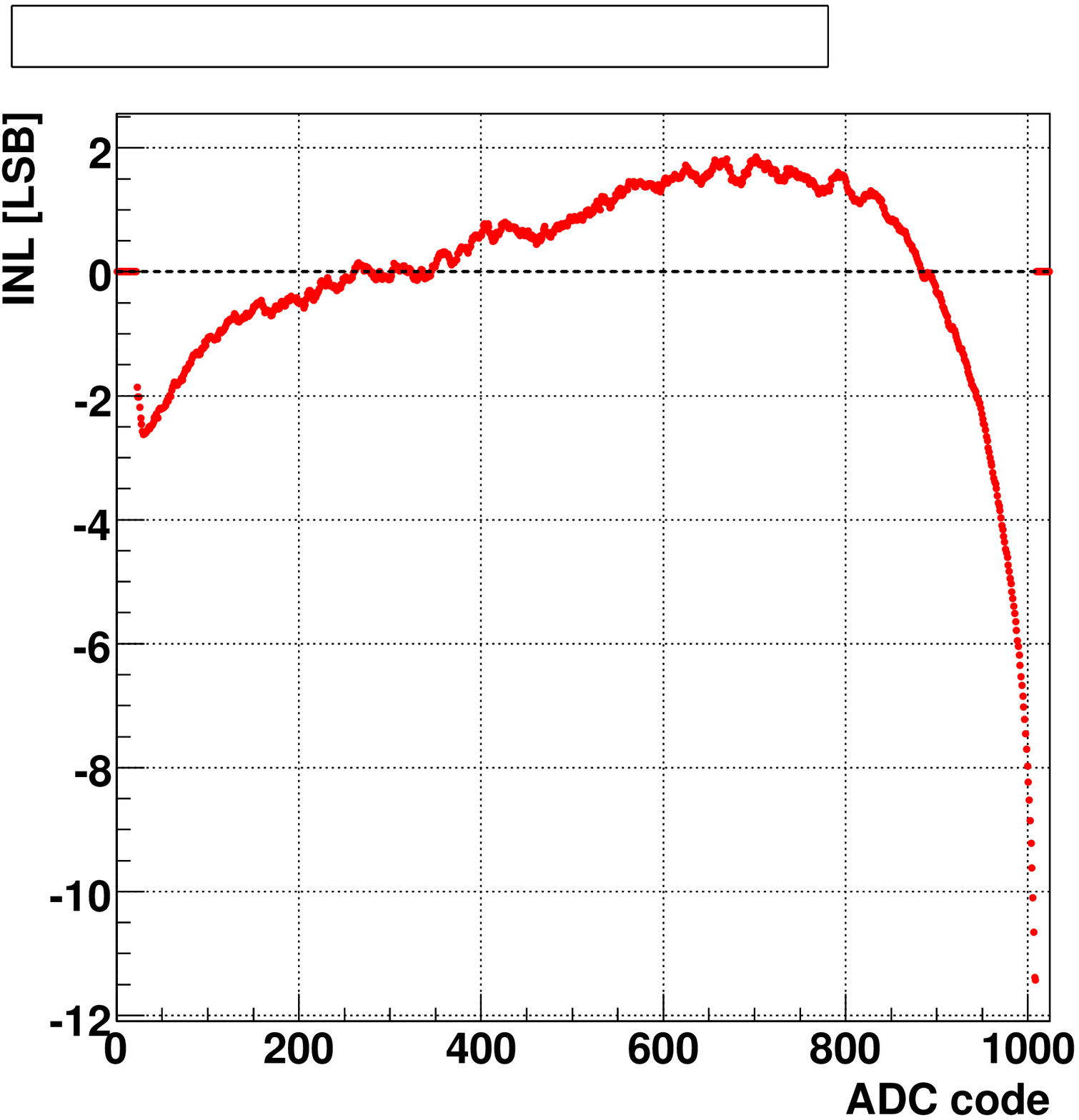, bb= 0 0 567 500, clip= , width=0.35\linewidth}}
    \put( 0.0,-0.1)  {(a)}
    \put( 4.8,-0.1)  {(b)}
    \put( 9.5,-0.1)  {(c)}
  \end{picture}
  \vspace*{-4.2mm}
  \caption{Linearity test: (a) measured and ideal prob.\ density function 
    of QDC transition codes, (b) differential non-linearity, (c) gain and 
    offset corrected integral non-linearity.}
  \label{fig:measurements}
\end{figure}

This test stand is based on CAMAC electronics and is used to develop different 
techniques for on- and off-line linearity measurements of various electronics 
components and different photodetectors. Up to now, two different methods for 
measuring the linearity of a QDC have been studied. The first method, of which 
Fig.~\ref{fig:ElectronicsTest} illustrates the setup, uses a sine wave as input 
to the QDC. The transition codes, or rather, the probability $P_{code}$ for 
each transition to occur at a certain ADC-code is measured and compared to the 
response of an ideal QDC, see Fig.~\ref{fig:measurements}(a):
\begin{displaymath}
  P_{code} = \frac{ N_{meas} }{ \pi \cdot \sqrt{\frac{A}{2}^2 - (code-offset)^2} },
\end{displaymath}
where $A=256$~pC is the charge amplitude at full scale range. 
Figure~\ref{fig:measurements}(b) shows the differential non-linearity 
(DNL), i.e., the difference between the measured and ideal QDC codes, 
while Fig.~\ref{fig:measurements}(c) displays the corresponding 
integral non-linearity (INL) after gain and offset correction, exemplary for 
channel 1. If a straight-line-fit is applied in the mid range of codes, from 200 
to 800 ADC-counts, the INL ranges from 1 to 2 LSBs (least significant bit, or 
ideal QDC code width), corresponding to 0.1-0.2\% of the full scale range. 

The second method is a QDC self-test, integrated in the readout software. 
A DC-voltage from 0 to 20~V is applied to the QDC (1~V $\widehat{=}$ 12.5~pC) 
and the measured charge is compared to the input charge for multiple charge 
injections. This second method is less precise and, moreover, requires many 
steps and thus a very long measurement time. 

In the near future, the component test stand will be used for a variety of 
other measurements, including the characterisation of different photo detectors 
(conventional PMTs and rather new developments, e.g. Silicon Photomultipliers), 
further linearity measurements regarding the photo detectors and, possibly, 
the entire readout chain. Longer term plans for this test stand also foresee 
measurements of temperature effects, the gain stability and other issues.

\section{The SLD Cherenkov Detector Test Stand}
\begin{figure}[!h]
  \setlength{\unitlength}{1.0cm}
  \begin{picture}(15.0, 10.0)
    \put( 0.2, 0.0)  {\epsfig{file=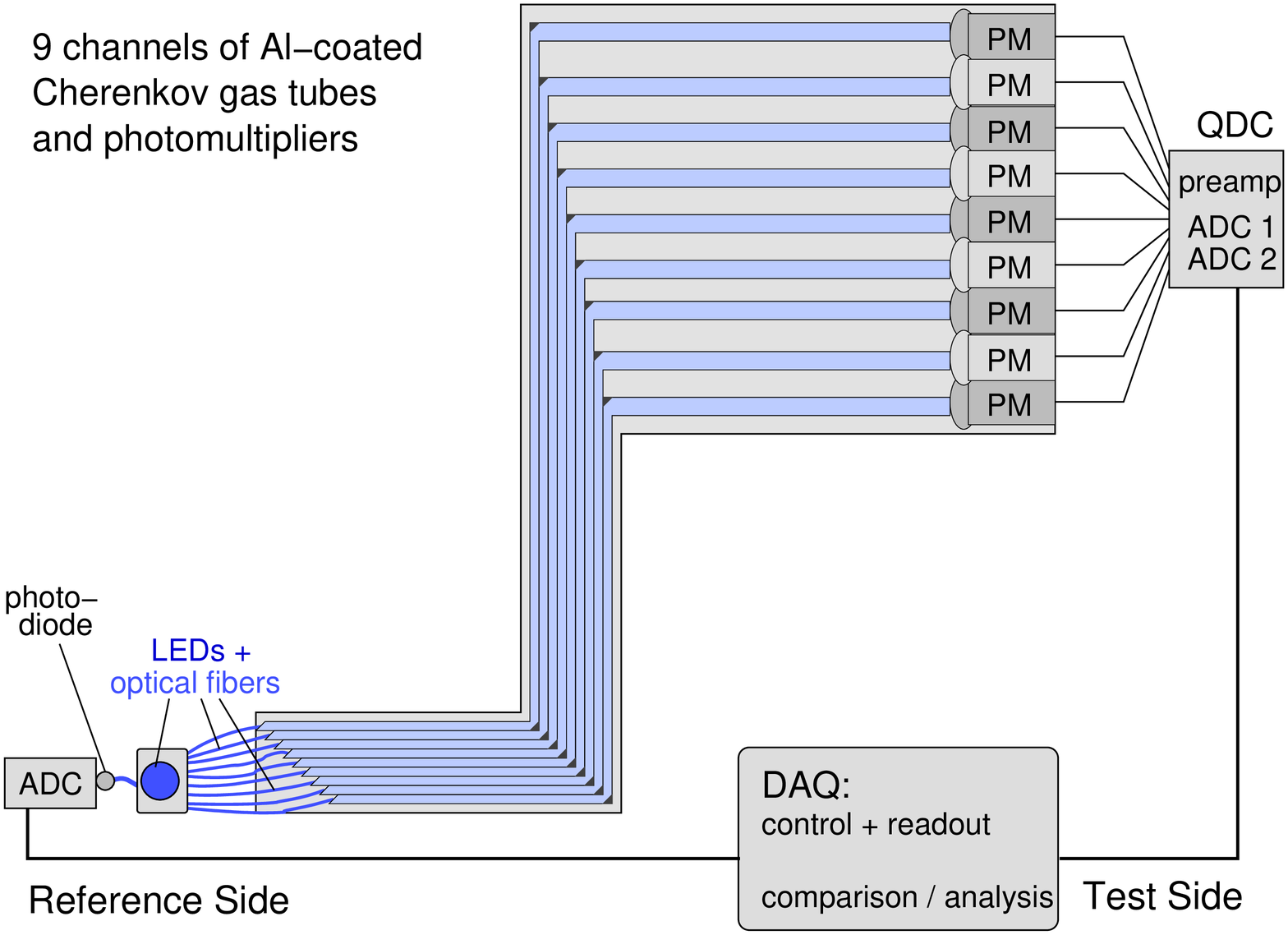,       bb= 0   0 800 580, clip= , width=1.01\linewidth}}
    \put(-0.1, 4.4)  {\epsfig{file=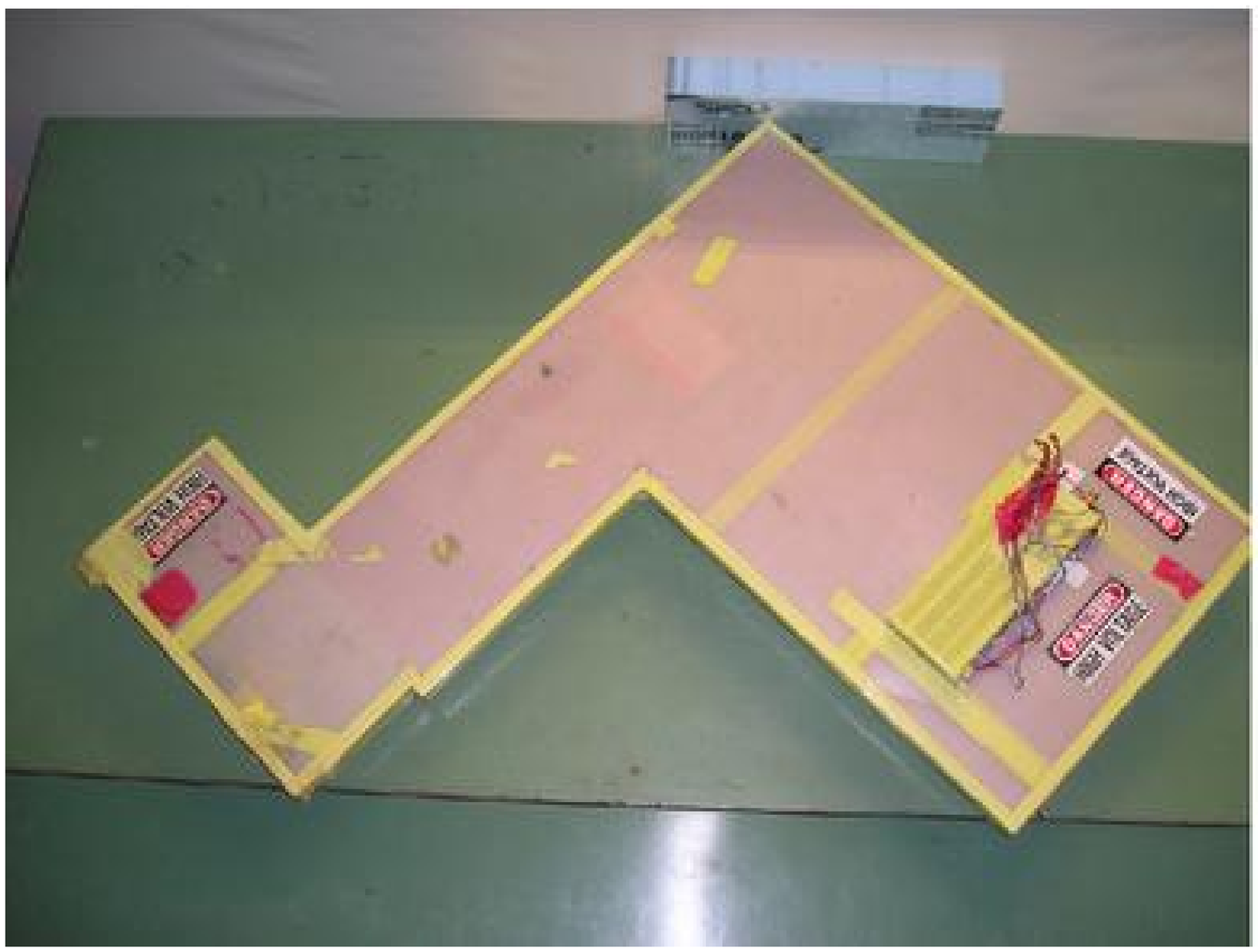, bb= 25 60 575 400, clip= , width=0.42\linewidth}}
    \put( 8.2, 2.3)  {\epsfig{file=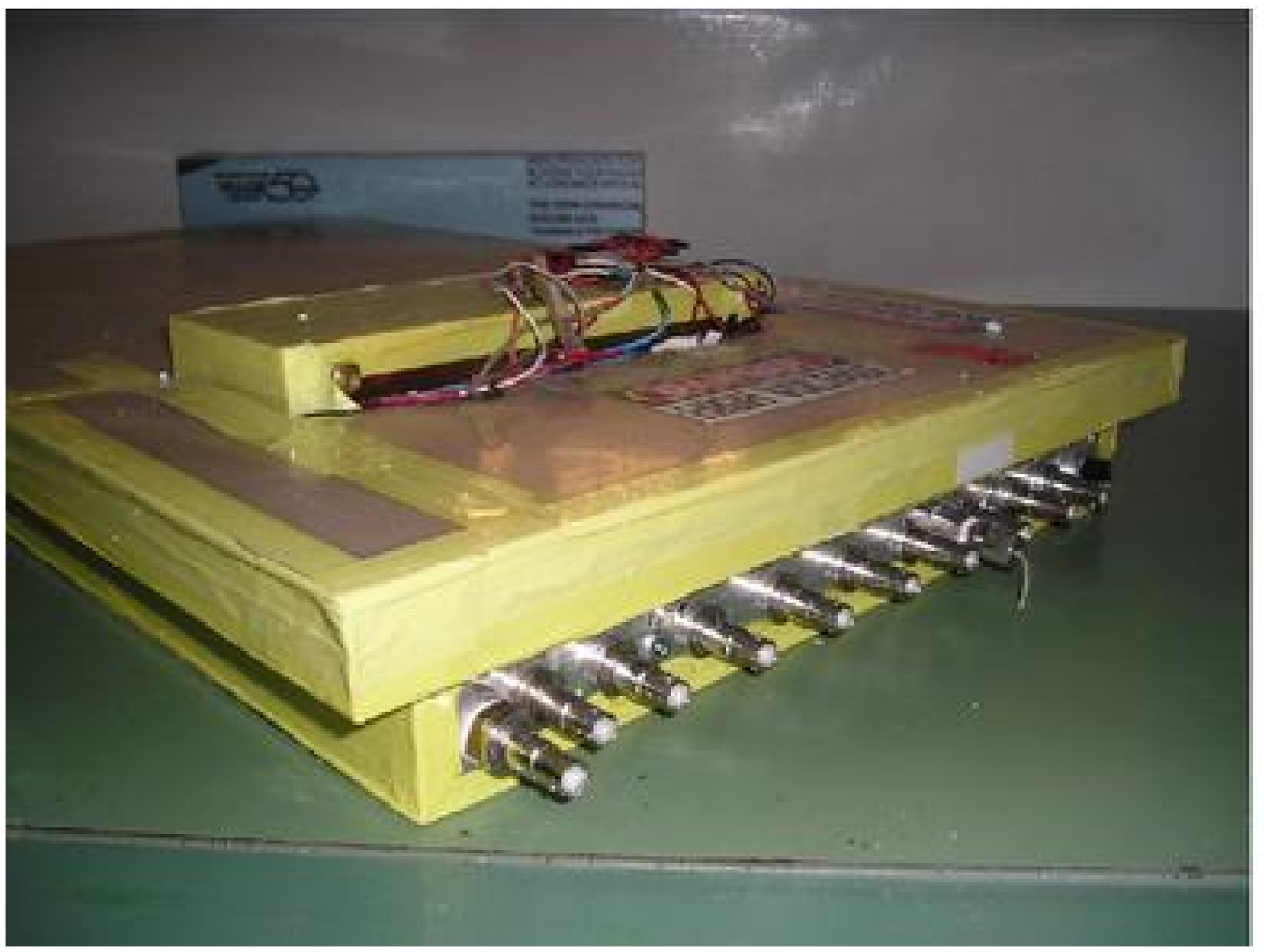, bb= 20 55 570 380, clip= , width=0.35\linewidth}}
  \end{picture}
  \caption{Schematic illustration of the setup of the old Cherenkov gas detector at DESY. 
    The two photographs show a birdseye view (left), and the frontend (right).}
  \label{fig:SLD-det}
\end{figure}

As of early May 2006, the SLD Cherenkov detector is located at DESY. However, 
the necessary hardware components for its setup are not yet available, but a 
VME-PCI interface as well as charge-sensitive ADC (QDC) have been ordered. 
(For a description of the entire SLD-detector and the Compton polarimeter 
see Ref.~\cite{bib:SLD-det}.)
For the detector commissioning it is planned to first test all nine channels 
for functionality -- with a system of blue (and green) LEDs. 
Furthermore, the reflectivity and light yield ($\leftrightarrow$ geometry), the 
sensitivity of the photo detectors, and the light extraction from the gas tubes 
will be studied. Later on, temperature effects will also be investigated, which 
might lead to an active regulation / stabilisation via thermo-electric elements. 
A proposal from April 18, 2007 lists further planned measurements, that will 
serve as a reference for studying new design features. Each measurement will 
either be performed as component and readout test of a single channel or of 
the entire detector system (Fig.~\ref{fig:SLD-det}), including:
\begin{itemize}
  \item  the characterisation of different types of photomultipliers (regarding sensitivity)
    \begin{itemize}
      \item[$\triangleright\;$]  dark rate / light response; 
      \item[$\triangleright\;$]  voltage and/or temperature dependence; 
      \item[$\triangleright\;$]  dynamic range / sensitivity;
    \end{itemize}
  \item  the Pros/Cons of different types of photo detectors and connecting fibers:
    \begin{itemize}
      \item[$\triangleright\;$]  photo detectors: conventional PMs, APDs, and SiPMs;
      \item[$\triangleright\;$]  fiber types: optical, wavelength-shifting fibers;
    \end{itemize}
  \item  and the analysis of different couplings: (direct, air gap, etc.)
    between gas tubes \& fibers, and between fibers \& photo detectors;
  \item  Linearity / non-linearity measurements for different configurations
\end{itemize}

\section{Acknowledgments}
The authors acknowledge the support by DFG Li 1560/1-1.


\begin{footnotesize}
%
%




\begin{thebibliography}{99}
\bibitem{url} Slides: \\ 
\verb$http://ilcagenda.linearcollider.org/materialDisplay.py?contribId=172&sessionId=78&material$
\verb$Id=slides&confId=1296$

\bibitem{bib:PhysicsCase}  
  G.A.~Moortgat-Pick and H.~Steiner, LC-TH-2000-055, 2000, \protect[\verb+http://www.desy.de/~lcnotes+\protect]; \\
  G.A.~Moortgat-Pick et al., ILC Reference Design Report, \protect[arXiv:hep-ph/0507011\protect].

\bibitem{bib:Tesla} TESLA Report 2001-23, Part III, DESY 2001-011, March 2001, 
  \protect[\verb+http://tesla.desy.de+\protect]; \\
  V.~Gharibyan, N.~Meyners, and K.P.~Sch\"uler, LC-DET-2001-047, DESY, Feb. 2001.

\bibitem{bib:SLD-det}
  R.D.~Elia, SLAC-Report-429, SLAC, Apr. 1994, (Ph.D.~Thesis); \\
  R.C.~King, SLAC-Report-452, SLAC, Sep. 1994, (Ph.D.~Thesis); \\
  The SLD Collaboration (M.~Woods), SLAC-PUB-7319, SLAC, Oct. 1996; \protect[arXiv:hep-ex/9611005\protect].

\end{thebibliography}
%

\end{footnotesize}


\end{document}